%% file: ipsn10-ETAP.tex
\documentclass{sig-alternate-ipsn10}

\usepackage{graphicx}
\usepackage{url}
\usepackage{amsmath,amssymb,amsfonts}
\usepackage{color}

\input{header.tex}

\numberofauthors{2}
\author{
\alignauthor Neal Patwari\\
\affaddr{University of Utah}\\
\affaddr{Salt Lake City, USA}\\
\email{npatwari@ece.utah.edu}
\alignauthor Joey Wilson\\
\affaddr{University of Utah}\\
\affaddr{Salt Lake City, USA}\\
\email{joey.wilson@utah.edu}
}

\title{People-Sensing Spatial Characteristics\\of RF Sensor Networks}

\begin{document}

\maketitle

\begin{abstract}
An ``RF sensor'' network can monitor RSS values on links in the network and perform device-free localization, \ie, locating a person or object moving in the area in which the network is deployed.  This paper provides a statistical model for the RSS variance as a function of the person's position w.r.t. the transmitter (TX) and receiver (RX).  We show that the ensemble mean of the RSS variance has an approximately linear relationship with the expected total affected power (ETAP).  We then use analysis to derive approximate expressions for the ETAP as a function of the person's position, for both scattering and reflection.  Counterintuitively, we show that reflection, not scattering, causes the RSS variance contours to be shaped like Cassini ovals.   Experimental tests reported here and in past literature are shown to validate the analysis.
\end{abstract}


\section{Introduction}

% I. Tag-free localization of people and objects using RF probing is possible, and useful.
% Define scope of research.  a) Multistatic, b) radio frequency (narrowband or wideband), c) device-free, d) model-based.

%In secure buildings, for example, active RFID systems are insufficient because they do not locate people who do not wear an active RFID tag.  During emergency building evacuations (\eg, during a fire), a DFL system would enable emergency responders (\eg, firefighters) to be aware of the locations of all people within the building.  

In this paper, we develop methods for the analysis of wireless sensor networks which perform device-free localization (DFL) of people and objects.  DFL networks locate people or objects without requiring them to cooperate in the system by carrying a radio device or a transmitting tag \cite{reggiani09,kanso09b,pratt2008dual,seifeldin2009nuzzer,zhang2007rf,wilson09a,wilson09b,wilson09c}.  Instead, DFL networks measure changes in the received signal strength (RSS) between many pairs of wireless sensors in the network to perform estimation of the locations of moving people and objects in the deployment area.  This is advantageous in applications like emergency response, or secure buildings, in which one cannot expect that people who must be located will be wearing a transmitter tag.

The DFL networks we study in this paper have five distinguising characteristics.  First, they are \textit{device-free}, in that the person (in this paper, we use ``person'' generally to refer to any large moving object) being tracked is not assumed to be wearing or attached to a transmitting or receiving radio device.  Second, sensor nodes are \textit{radio frequency (RF)} sensors.  RF has particular advantages of being able to penetrate (non-metal) walls and smoke.  Thirdly, they are sensor networks, in particular, peer-to-peer networks of RF sensors, and thus measurements can be made between many pairs of RF sensors.  Fourthly, they are model-based, \ie, sensor locations are known, but there is no training nor \textit{a priori} environment knowledge.  Finally, they are \textit{narrowband}. It is also possible to use ultra wide-band signalling to perform DFL \cite{chang04,rydstrom08,reggiani09}.  This paper focuses on less expensive narrowband sensors, which cannot distinguish multipath which arrive with different time delays.  

%Device-free localization does not provide identification.  We may know where a person is, but we cannot identify this person without other methods.  Privacy concerns are certainly important in device-free localization, but are less significant than in video surveillance, which can additionally be used to identify a person.

% RF-based device-free localization may be either wideband or narrowband.  If radios in the system transmit and receive wideband pulses, such as ultra-wide band (UWB) pulses, the sensing system has the ability to distinguish multipath components at different time delays.  Systems with narrowband radio sensors simply measure the received power of the transmitted signal.  In theory, sensors could measure the phase of the received narrowband signal, however, this would require phase-coherence between distributed sensors and thus we do not consider this possibility. Sensors which measure received power are the focus of this paper, although the models and methods presented could be used to analyze the phase-coherent and wideband cases.

% II. Physics paragraphs

Variance of RSS due to human or object motion is due to changes in individual multipath which contribute to the received signal.  The complex baseband voltage for a narrowband signal measured at a receiver is \cite{durgin02},
\begin{equation}\label{E:voltageReceiver}
 \tilde{V} = \sum_{i=1}^N V_i,
\end{equation}
where $V_i = |V_i| e^{j \phi_i}$ is the complex amplitude of the $i$th multipath component, and $N$ is the total number of components.  The received power is equal to $| \tilde{V}|^2$, the squared magnitude of the complex baseband voltage.  What we call the ``received signal strength'' (RSS) is actually the received power in decibel terms, that is,
\begin{equation}
 R_{dB} = 20 \log_{10} \| \tilde{V}\| \nn
\end{equation}
% Each multipath component travels a different trajectory from transmitter to receiver.  We denote the subset of space through which multipath $i$ travels as set $S_i$.  Using a plane wave approximation, we approximate each trajectory as a sequence of connected line segments from the transmitter to the receiver.

%In small-scale fading, as an antenna moves, the phases $\{\phi_i\}_i$ change at different rates for different $i$, and thus the magnitude of the phasor sum in (\ref{E:phasorSum}) changes.  
Fading is the effect of the changes in phases $\{\phi_i\}_i$.  Depending on these phases, the sum in (\ref{E:voltageReceiver}) may be destructive (opposite phases) or constructive (similar phases).  When objects and people in the environment move, they affect a subset of the multipath components depending on their current position.  The new person or object may also scatter radio power and thus induce a new multipath component in the channel. In the bistatic radar literature, this new scatter-path is the only channel change considered \cite{paolini08}.  In this paper, we provide new models which allow us to quantify the total fading caused by changes \textit{in existing} multipath components, as a function of the position of the new person.

% \begin{figure}[htbp]
%   \centerline{ (a) \psfig{figure=example_without_obj.eps,width=2in}}
%   \centerline{ (b) \psfig{figure=example_with_obj.eps,width=2in}}
%   \caption{(a) Example spatial extent of multipath between the transmitter at $\mbx_t$ and the receiver at $\mbx_r$.  (b) When a person appears at $\mbx_o$, it can cause additional scattering paths (\textcolor{red}{----}) and can alter previously existing multipath (\textcolor{blue}{-----}).}
%   \label{F:EgAddChangeMultipath}
% \end{figure}

%Since object position-specific relationships had not been important in either communications or radar applications, such models have not, in the past, been explored in the literature.

Fading due to human motion has been quantified in past studies in order to aid in the design of communications systems which operate in such fading environments \cite{bultitude87}.  Researchers have observed that the motion of people near either a transmitter or receiver impacts measured fading statistics, and when more people are moving in proximity of a transmitter or receiver, fading increases \cite{hashemi94}.  

\begin{figure}[htbp]
  \centerline{ \psfig{figure=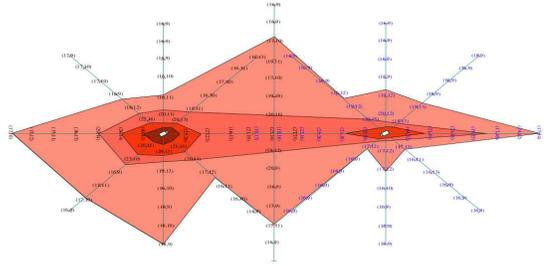,width=2.8in}}
%  \caption{From \cite{yao2008model}, fading signal ``variation'' as a function of object position.}
  \caption{From \protect\cite{yao2008model}, fading signal ``variation'' as a function of a person's position.}
  \label{F:yao-gao-figure4}
\end{figure}

Of particular interest for DFL, experimental results have been presented which relate a person's coordinate (with respect to TX and RX coordinates) to the variance measured on a link. In \cite{yao2008model}, a contour plot, reproduced in Figure \ref{F:yao-gao-figure4}, shows that the ``variation'' (sum of absolute values of the differences of the time series) is highest when a person is located close to either node, is relatively high when a person is in the line segment between the nodes, and decays away from this line and the nodes.  In contrast, experimental results in \cite{zhang2007rf} show a different result, that the ``dynamic'' (average absolute value of the difference from the static mean) is highest in an oval centered at the midpoint of the line segment between the two nodes.  We show in this paper how both results can be explained using analysis.  We also present in this paper results of a new experiment conducted in our University bookstore which also validates the analysis.

% II. Goal and Methods
%   A. Goal: explain the relationship between the severity of channel fading at a receiver, to the position of a person or object (wrt the tx and rx).
%   B. Methods: via traditional spatial multipath models, including scattering and reflection propagation mechanisms.

The goal of this paper is to provide a statistical model which relates (1) the variance of channel fading measured on a link to (2) the position of the moving person in the environment w.r.t.~the TX and RX locations.  We desire a model which is based in the physics of radio wave propagation and does not require specific knowledge of the environment.  For these purposes we apply traditional statistical spatial propagation models for scattering and reflection, which were originally developed to study multi-antenna or directional antenna systems \cite{liberti96,norklit98}.

% III.  Summary of results
%   A. The number of multipath 

We obtain the statistical model in three steps.  First, in Section \ref{S:Variance_ETAP}, we show that the ensemble mean (across many realizations of a link) of RSS variance is approximately linearly related to the expected total affected power (ETAP) in dB.  Secondly, in Section \ref{S:MultipathModels}, we present statistical multipath models for scattering and reflection.   Finally, in Section \ref{S:Analysis}, we quantify the ETAP a function of the person's position (with respect to the TX and RX positions).  

A counter-intuitive result of the analysis is that RSS variance has contours similar to Cassini ovals, as would be seen in a bistatic \textit{scattering} power equation; however, this dependence occurs when the sole multipath propagation mechanism is \textit{reflection}.  We verify the new model using measurements in Section \ref{S:ExperimentalResults} -- those reported in the literature \cite{yao2008model,zhang2007rf}, and an extensive new measurement campaign we conduct and present in this paper.

\section{Related Work}

% III. Related research of RF-based multistatic probing
%   A. UWB-based
%   B. UNB
%   C. Many CS approaches

In the Introduction, we reviewed propagation research which reported the effects of human motion on fading statistics.  DFL takes advantage of advances in wireless sensor networks to develop networks of ``RF sensors'' for which the radio is the sensor.  Woyach, Puccinelli, and Haenggi \cite{woyach06} were the first to present RSS-based DFL, to our knowledge, and termed these applications as ``sensorless sensing''.  As one example, they showed how RSS measurements from sensors deployed in a hallway could be used to track someone walking through, and to indicate whether they were walking or running.  Zhang et.~al.~\cite{zhang2007rf,zhang2009dynamic} positioned sensors on a ceiling and used the change in RSS on links to estimate the location of people moving below.   

Youssef, Mah, and Agrawala \cite{youssef07} coined the term ``device-free passive (DfP) localization'' to distinguish this system, in which people are located regardless of whether they carry a radio, from active RFID systems which locate a TX or allow a RX to locate itself.  Two detection methods and a radio map-based tracking method are presented in \cite{youssef07}.  Other work has also used training measurements to perform localization and tracking \cite{viani2008object,seifeldin2009nuzzer,moussa2009smart}.  Because training measurements require significant time, automatic generation of radio maps is ``one of the main technical challenges behind the implementation of DfP'' and it is critical to know ``the effect of these entities [in the environment] on the received signal at different points in the environment'' \cite{youssef07}.  This paper provides statistical models and tools for this purpose.

Our past research has also used changes in the statistics (mean and variance) of RSS in order to estimate an image of presence of people in the environment \cite{wilson09a,wilson09b,wilson09c}.  In particular, we have shown the ability of an RF sensor network to track a person moving inside of a building through external walls \cite{wilson09c}.  The models developed in this paper are useful for development of estimation bounds and improved estimation algorithms.

Other related efforts have explored details regarding the effect of human movement on measured RSS.  An experimental test in \cite{lee2009wireless} uses RSS on a link to reliably detect the presence of a single person walking from one node to the other, in a variety of different indoor and outdoor environments.  Experiments in \cite{nakatsuka08} go further by providing  linear relationships between the RSS mean and variance and the number of people walking or sitting between two nodes, for use in crowd density estimation.  Very relevant to this paper, measurements reported in  \cite{yao2008model} show the ``variation'' of RSS caused by human movement as a function of position across the 2D plane containing the two nodes.  

Research has also investigated polarization \cite{pratt2008dual} as an alternative to RSS.  When two antennas can be arranged to have orthogonal polarization, it adds significantly to the ability of a detector to measure a change in the environment \cite{pratt2008dual}.  

Further, ultra wide-band (UWB) measurements have long been suggested for through-wall surveillance (TWS).  Multi-static UWB approaches use networks of UWB transceivers to locate the position of a new person in the environment within and around the network and thus are closely related to this research \cite{chang04,paolini08,reggiani09,rydstrom08}.  From the radar perspective, a new person in the environment creates an additional scattering multipath component.  In multipath-rich environments, such as indoor environments, the changes to existing multipath are at least as important as any additional scatter-path caused by the presence of a new person. This paper focuses on the effects of a new person on the existing multipath in the channel.

\section{Relation of RSS Variance and ETAP} \label{S:Variance_ETAP}

In this section we show that the variance of RSS has an approximately linear relationship with the expected total affected power (ETAP), that is, the expected value of the powers in affected multipath components.  In other words, a person's new position causes random changes in some of the multipath components, and the sum of the power in these affected components has a linear relationship with the variance of fading.  We start by showing how random changes in affected multipath, and no changes in other multipath, lead to a Ricean distribution of fading.  We present a linear relationship between the variance of RSS (in dB) and the K factor (in dB) of the Ricean distribution.  Then we show that the ensemble mean of the K factor (in dB) has an approximately linear relationship with the ETAP.

To define ETAP, consider the changes in the complex baseband voltage in (\ref{E:voltageReceiver}) that occur when a moving person now exists at position $\mbx_o$.  We assume that some of the multipath components $i$ are affected in phase and amplitude.  Other multipath components, because their path does not come close to the person, are unaffected by the change.  Define sets $\mathcal{T}$ and $\mathcal{T}'$ as the indices of multipath which are unaffected, and affected by the new person, respectively.  ETAP is defined as  
\begin{equation}\label{E:ETAP}
 \mbox{ETAP} = \E{}{\sum_{i\in \mathcal{T}'} |V_i |^2}
\end{equation}
where $V_i$ is the complex baseband voltage of multipath $i$.  ETAP is a function of that person's position $\mbx_o$, as will be discussed extensively in Section \ref{S:Analysis}, but first, we show its relationship with RSS variance.

\subsection{Ricean Distribution of Fading}

As discussed above, we can consider multipath to be either affected or unaffected by the moving person at position $\mbx_o$.  Consider partitioning the sum in (\ref{E:voltageReceiver}) into two sums,
\begin{equation}\label{E:voltageReceiverGrouped}
 \tilde{V} = \sum_{i \in \mathcal{T}} V_i + \sum_{i\in \mathcal{T}'} V_i 
\end{equation}
The first sum over $i \in \mathcal{T}$ is unaffected by the placement of the new person, and therefore is constant (\ie, non-random) for a particular link.  We denote this first sum as $\bar{V} = \sum_{i \in \mathcal{T}} V_i$.  The second sum over $i\notin \mathcal{T}$ is random because the multipath components change randomly because of the motion of the person.  We assume that the phases are randomly altered (uniform between 0 and $\pi$) and multipath components $i \notin \mathcal{T}$ are multiple. Applying the central limit theorem, the real and imaginary parts of the sum of the affected multipath complex baseband voltages are seen to be independent and identically distributed (i.i.d.) zero-mean Gaussian random variables $V_{ns,I}$ and $V_{ns,Q}$ \cite{durgin02}.  So,
\begin{equation}\label{E:voltageReceiverRandom}
 \tilde{V} = \bar{V}  + V_{ns,I} + j V_{ns,Q}
\end{equation}
The variance of $V_{ns} = V_{ns,I} + j V_{ns,Q}$ is equal to the sum of the magnitude squared of the voltages in its sum, that is, $\sum_{i\notin \mathcal{T}} |V_i |^2$ \cite{rappaport96}.

The envelope $\| \tilde{V}\|$ is Ricean because it is the magnitude of a complex Gaussian random variable with non-zero mean $\bar{V}$.  If there were no unaffected multipath, \ie, $|\mathcal{T}|=0$, then the envelope would reduce to a Rayleigh random variable.  The $K$-factor of this Ricean distribution is given by
\begin{equation} \label{E:K}
 K  = \frac{|\bar{V}|^2}{\sum_{i\in \mathcal{T}'} |V_i |^2}
\end{equation}

\subsection{Relation of RSS Variance and K Factor}

In past research, we have shown that the variance of $R_{dB}$, when the envelope is Ricean, is purely a function of the Ricean K-factor of the envelope \cite{wilson09c}.  In fact, there is an approximately linear relationship between the variance of $R_{dB}$, $\Var{}{R_{dB}}$, and the K-factor in decibels, $K_{dB} = 10 \log_{10} K$, for a range of $-2 < K_{dB} < 10 dB$ \cite{wilson09c}.  In this linear region, $\Var{}{R_{dB}}$ varies from 27 dB$^2$ to 3 dB$^2$.  This linear region is when the variance of $R_{dB}$ is most important.  Stationary links will measure $\Var{}{R_{dB}} < 3 \mbox{ dB}^2$ due to noise.  In our experience, very few links with TX and RX stationary will measure $\Var{}{R_{dB}} > 27 \mbox{ dB}^2$.  Thus, for our purposes, it is accurate to describe the relationship between $\Var{}{R_{dB}}$ and $K_{dB}$ as linear.  Furthermore, it is decreasing - as $K_{dB}$ increases, it means less severe fading is occurring, and thus a lower $\Var{}{R_{dB}}$.  To be explicit, 
\begin{equation} \label{E:linear_R_dB}
\Var{}{R_{dB}} \approx a_0 - a_1 K_{dB},
\end{equation}
for some positive constants $a_0, a_1$.

In dB terms, (\ref{E:K}) becomes
\begin{equation} 
 K_{dB}  = 10\log_{10} |\bar{V}|^2  - 10\log_{10} (\sum_{i\in \mathcal{T}'} |V_i |^2) \nn
\end{equation}
Effectively, $K_{dB}$ is the power in the sum of unaffected multipath minus the power in the affected multipath.  What is the expected value of $K_{dB}$, taken across an ensemble of environments with the same geometry of TX, RX, and person position?
\begin{equation} \label{E:E_K_dB}
 \E{}{K_{dB}}  = \E{}{10\log_{10} |\bar{V}|^2}  - \E{}{10\log_{10} \sum_{i\notin \mathcal{T}} |V_i |^2}
\end{equation}
In relative terms, the power in the unaffected multipath are generally much stronger than the power in the affected multipath.  With multipath dispersing in all directions around an environment, a person typically affects only some small fraction of the total power (unless the person is right on top of the TX or RX).  If we rewrite the left term, $\E{}{10\log_{10} |\bar{V}|^2}$, as the expected value of the $10\log_{10}$ of (total power - affected multipath power), we can see that it is largely constant when the affected multipath power is much less than the total power.  We approximate (\ref{E:E_K_dB}) as,
\begin{equation} \label{E:E_K_dB_approx}
 \E{}{K_{dB}}  \approx c_{up}  - \E{}{10\log_{10} \sum_{i\notin \mathcal{T}} |V_i |^2}
\end{equation}
for a constant $c_{up}$.  When the affected power is one half of the total power, this constant approximation is in error by 3 dB. When the affected power is less, the error is respectively less.

Combining (\ref{E:E_K_dB_approx}) and (\ref{E:linear_R_dB}), we have that 
\begin{eqnarray}
 \E{}{\Var{}{R_{dB}} } &\approx& a_2  + a_1 \E{}{10 \log_{10} \sum_{i\in \mathcal{T}'} |V_i |^2} 
\nn
\end{eqnarray}
for $a_2 = a_0 - a_1 c_{up}$.  Note again that the variance is taken in one particular environment for one particular link, while the expected value is across the ensemble of environments and links with the same geometry.  

Finally, we approximate $\E{}{\log Y} \approx \log \E{}{Y}$, for $Y = \sum_{i\in \mathcal{T}'} |V_i |^2$.  We show in Appendix \ref{S:ApproxExpectedLog} that this is a good approximation when $|\mathcal{T}'|$ is more than one or two, \ie, multiple multipath are affected by the person's position.  With this approximation,
\begin{equation} \label{E:Final_E_Var_R_dB}
 \E{}{\Var{}{R_{dB}} } \approx a_2  + a_1 10 \log_{10} \E{}{\sum_{i\notin \mathcal{T}} |V_i |^2} 
\end{equation}

In summary, the ensemble mean of RSS variance has an affine relationship with the expected value of the sum of the powers in affected multipath components, \ie, ETAP.  Sections \ref{S:MultipathModels} and \ref{S:Analysis} analyze ETAP, in particular, how it is a function of the propagation mechanisms and TX, RX, and person's positions.

\section{Statistical Multipath Models} \label{S:MultipathModels}

To quantify fading as a function of an person's position in space, we need a statistical channel model which describes the spatial extent of each multipath in the channel, what we term a \textit{spatial multipath model}.  Several spatial multipath models have been developed in past research for the study of directional antennas in cellular systems \cite{liberti96,norklit98}, for the development of fading models \cite{durgin}, and for multiple-input multiple-output (MIMO) systems \cite{kermoal02}, among other purposes \cite{ertel98}.  We focus on models which explain the geometric path of each multipath component, rather than those based purely on measurements which model only angle-of-arrival (AOA) or time-of-arrival (TOA) at the receiver.  We also need models applicable to the case where antenna heights are both relatively close to the ground (as opposed to being at cellular base station heights).

\subsection{Related Work}

%, the GBSBM allows determination of the joint probability density of AOA and TOA of the channel's multipath components \cite{liberti96}

The geometrically-based single-bounce model (GBSBM) \cite{liberti96} is a spatial multipath model used in cases when the antenna is at the same height as the scatterers.  In this model,  each (non-LOS) multipath experiences a single reflection.  Any multipath component which arrives with excess delay less than a threshold is known to have changed direction at some point within an ellipse.  These reflection points are assumed to be uniformly distributed in the environment.  In \cite{liberti96}, the received power at location $\mbx_r$ of the multipath component originating at $\mbx_t$ and reflected at $\mbx$ is given by,
\begin{equation} \label{E:Liberti2}
 P_r(\mbx) = \frac{c_r}{( \| \mbx_t - \mbx \| + \| \mbx_r - \mbx \| )^{n_p}}
\end{equation}
where $c_r$ is a constant, and $n_p$ is the path loss exponent.

The model of \cite{norklit98} is also a geometrical model which assumes each path is scattered once by a scatterer on a plane.  In \cite{norklit98}, however, the plane does not contain the TX and RX antennas, which may exist at arbitrary heights $h_{BS}$ and $h_{MS}$ for the base station and mobile station, respectively.  The model of \cite{norklit98} is also single-bounce, but the propagation mechanism is scattering, so that the received power of a scattered multipath is given by,
\begin{equation} \label{E:Norklit}
 P_s(\mbx) = \frac{c_s}{\| \mbx_t - \mbx\|^2 \| \mbx_r - \mbx\|^2}
\end{equation}
where $c_s$ is a constant.

\begin{figure}[htbp]
  \centerline{ \psfig{figure=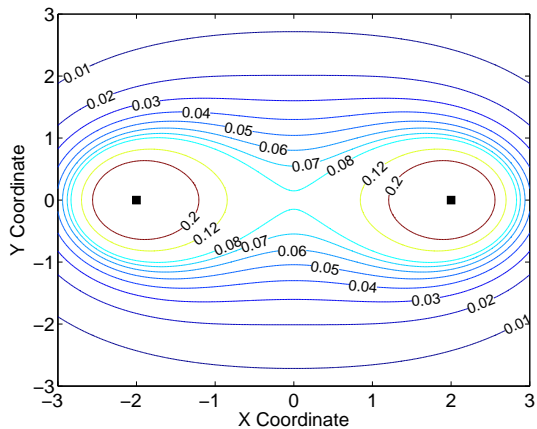,width=2.8in}}
  \caption{A contour plot of Cassini ovals for the case when TX and RX locations are shown as $\blacksquare$, and $c_s=100$.}
  \label{F:plot_Cassinis_ovals}
\end{figure}

The locus of points $\{\mbx: P_s(\mbx) = \gamma \}$ for some constant $\gamma$ is called a \textit{Cassini oval} with foci at $\mbx_t$ and $\mbx_r$ \cite{paolini08}.  For different values of $\gamma$, Cassini ovals are plotted in Figure \ref{F:plot_Cassinis_ovals}.  In general, $P_s(\mbx)$ is low when the scatterer at $\mbx$ is far from both the TX and RX, and is high when $\mbx$ is close to one of them or the line in between them.

\subsection{Our Model} \label{S:OurModel}

For the statistical multipath model in this paper, we consider the union of the models of \cite{liberti96} and \cite{norklit98}.  To be specific, we make four simplifying assumptions about the radio channels in the environment of interest:
\begin{enumerate}
 \item Omni-directional antennas are used at each node.
 \item Mechanisms for propagation include scattering and reflection. Diffraction is not considered in this paper.
 \item The scatterers (and reflectors) are located in a plane parallel to the ground, the ``scatterer plane''. 
 \item Single-bounce propagation.  Each multipath has a single change in direction on its path from TX to RX.
\end{enumerate}
As they were in \cite{liberti96} and \cite{norklit98}, these are significant simplifying assumptions.  The single-bounce model is, as Chang and Sahai concisely described, ``a bit dubious'' \cite{chang04}, but it can help to provide a first-order model for multipath-induced characteristics.    More general analysis which assumes multiple reflections and directional antennas can be an extension of techniques described in this paper.   Diffraction loss has a complicated relationship with object geometry and angles of arrival and departure from the object, which makes general analysis difficult.

When we consider multipath which have experienced a single scatter at location $\mbx$, we describe their received power as $P_s(\mbx)$ as given in (\ref{E:Norklit}).  When we consider multipath which have experienced a single reflection at location $\mbx$, we describe their received power as $P_r(\mbx)$ as given in (\ref{E:Liberti2}).  To keep notation consistent, we describe the physical item which exists at location $\mbx$ as a ``scatterer'' regardless of whether the item reflects or scatterers the radio wave.  Each different scatterer might have a different constant $c_s$ or $c_r$, however, so for simplicity, we assume that both are constant across the ensemble of scatterers.  We have assumed isotropy of antennas and radar cross-section (RCS) and thus $c_s$ and $c_r$ are not a function of position.

We also note that the power of a scatter-path from the new person can also be described using our model. In past research \cite{paolini08,zhang2007rf,zhang2009dynamic} it is assumed that a new person in the environment creates an additional multipath component based solely on scattering from the person.  The received power of this new scatter-path from the new person at location $\mbx_o$ can be written as $P_s(\mbx_o)$.

Similar to \cite{liberti96,norklit98} we assume scatterer locations are Poisson distributed across the scatterer plane.  W.l.o.g, the scatterer plane z-coordinate is taken to be zero.  We denote the average scatterer density as $\eta$/m$^2$.   The plane containing all RF sensors is also parallel to the ground, \ie, the TX and RX locations are always at the same height, which we denote $\Delta z$.  

\begin{figure}[htbp]
  \centerline{ \psfig{figure=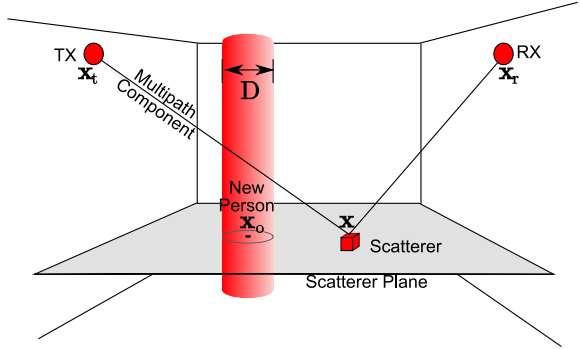,width=3.0in}}
  \caption{TX, RX, plane containing scatterers, and the new person.}
  \label{F:scatterer_object}
\end{figure}

We also assume that the person can be approximated by a vertical cylinder as shown in Figure \ref{F:scatterer_object}.  In the scatterer plane, it is a circle centered at $\mbx_o$ with diameter $D$.  The top of the cylinder is higher than the highest, and the bottom of the cylinder is lower than the lowest, of $\Delta z$ and zero.  This is only a crude approximation of a person's shape, but is simple to analyze because the cylinder casts an infinite ``shadow'' onto the scatterer plane.  Given a scatterer at location $\mbx$, if either the line segment connecting the TX to $\mbx$, or the line segment from $\mbx$ to the RX, intersects with this vertical cylinder, then the multipath is ``affected''.

\section{Analysis} \label{S:Analysis}

Given the model assumptions described in Section \ref{S:OurModel}, we analyze two types of ETAP, first assuming the propagation mechanism is scattering, and then assuming it is reflection.  Certainly, any real-world link will have multipath caused by both reflection and scattering.  In addition, there may be a line-of-sight path which is not reflected or scattered.  The real-world ETAP will be some linear combination of the ETAP due to scattering and the ETAP due to reflection.

\begin{figure}[htbp]
  \centerline{ \psfig{figure=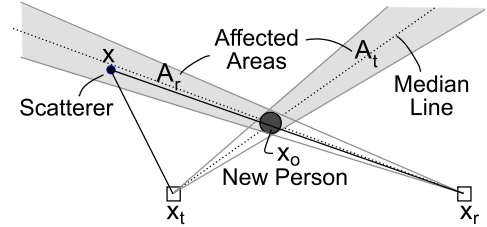,width=2.5in}}
  \caption{The TX $\mbx_t$, RX $\mbx_r$, and the ``affected'' area (shaded) in which a scatter could be located in order to have a single-bounce path that crosses the person located at $\mbx_o$.}
  \label{F:affectedAreasInEllipse}
\end{figure}

Key to this analysis is to see that single-bounce multipath components affected by a person have scatterers in the \textit{shadow} of that person, with respect to either the TX or RX.  By ``shadow'' with respect to the TX or RX, we mean that the line segment from either the TX or RX to any point in the shadow crosses through the person.  Figure \ref{F:affectedAreasInEllipse} shows a multipath component with a scatterer in the shadow of the person with respect to the RX.  The analysis of affected multipath transforms into an analysis of the positions of scatterers within an area $A = A_t \cup A_r$, where we denote the areas shadowed by the person with respect to the TX or RX as $A_t$ and $A_r$, respectively.

The area $A_t$ is essentially an infinite height isosceles trapezoid; equivalently, an infinite height isosceles triangle with the area between the vertex $\mbx_t$ and the person at $\mbx_o$ removed.  Figure \ref{F:affectedAreasInEllipse} shows the \textit{median line}, \ie, the line from $\mbx_t$ which passes through $\mbx_o$.  We make two simplifying assumptions:
\begin{itemize}
 \item \textit{Approximation 1}: The width of the shadowed area at $\mbx_o$ is equal to $D$, the diameter of the person.
 \item \textit{Approximation 2}: The value of a function $f(\mbx)$ for $\mbx$ in the shadowed area can be approximated by $f(\tilde{\mbx})$, where $\tilde{\mbx}$ is the projection of $\mbx$ onto the median line, \ie, the point on the median line closest to $\mbx$.  In other words, the isosceles trapezoid of the shadowed area is narrow enough that the value of $f(\mbx)$ can be assumed to be nearly constant along the perpendicular direction from the median line.
\end{itemize}
Both approximations are good when $D \ll \| \mbx_o - \mbx_t\|$.  These approximations are also used with $A_r$, the shadowed area by the person with respect to the receiver, by replacing $\mbx_t$ with $\mbx_r$ in the above description.

%We will compare exact numerical solutions with those obtained using these approximations in order to validate our analysis methods.  

In a homogeneous Poisson spatial scatterer process, scatterer locations are randomly distributed across the scatterer plane with density $\eta$ scatterers per unit area.  In each ETAP (scattering or reflection), we want to find the expected value of a function of the scatterer locations.  The ETAP is denoted $Q$ and is written as,
\begin{equation}
 Q = \E{}{\sum_{i} f(\mbx_i)} \nn
\end{equation}
where the sum is over every scatterer location, and where $f(\mbx_i)$ is a scalar function of scatterer position $\mbx_i$.  In general, we write this expected value as,
\begin{equation} \label{E:ExactIntegral}
 Q =  \iint_{\mbx \in A} \eta f(\mbx ) d\mbx.
\end{equation}
At this point, we make another approximation:
\begin{itemize}
 \item \textit{Approximation 3}:  We approximate the area integral in (\ref{E:ExactIntegral}) by the sum of two shadowed area integrals, one over $A_r$ and the other over $A_t$.  With this approximation,  $Q =  Q_t + Q_r$, where
\begin{equation} \label{E:Integral_approx3}
 Q_t =  \iint_{\mbx \in A_t} \eta f(\mbx ) d \mbx , \quad Q_r = \iint_{\mbx \in A_r} \eta f(\mbx ) d \mbx
\end{equation}
Essentially, we assume that the two parts of the shadowed area, $A_r$ and $A_t$, are nearly disjoint.
\end{itemize}
This approximation is good except for when $\mbx_o$ is on the ``far'' parts of the line containing $\mbx_t$ and $\mbx_r$, \ie, on the straight line containing $\mbx_t$ and $\mbx_r$ but not in between $\mbx_t$ and $\mbx_r$.  On these far parts of the line, the shadows of the person w.r.t.~the TX and RX overlap, and we do not consider the resulting expression to be accurate.

Because of approximations 1 and 2, the area integrals in (\ref{E:Integral_approx3}) can be simplified to be written as line integrals as follows.  First, consider the area shadowed w.r.t.~the TX, $A_t$.  Because of approximation 2, the value of $f(\mbx)$ is constant along the line segment perpendicular to the median line within the shadowed area.  To simplify, we write $\tilde{\mbx}$ as the projection of $\mbx$ onto the median line.  The vector $\tilde{\mbx}$ can be written as,
\begin{equation}
 \tilde{\mbx} = \mbx_o + \alpha \frac{\mbx_o - \mbx_t}{\| \mbx_o - \mbx_t \|}, \nn
\end{equation}
for some $\alpha \ge 0$.  The fraction $(\mbx_o - \mbx_t) / \| \mbx_o - \mbx_t \|$ is the unit vector parallel to the median line.  The width of the shadowed area at $\tilde{\mbx}$ is the length of the line segment perpendicular to the median line contained with $A_t$.  This width is calculated using similar triangles.  The isosceles triangle that starts at $\mbx_t$ and ends at the person location $\mbx_o$ has height $\| \mbx_t - \mbx_o\|$ and has base length approximately equal to $D$ (by approximation 1).  The full isosceles triangle has height $\| \mbx_t - \mbx_o\| + \alpha$, and we denote its base length as $b_t$.  So,
\begin{equation} 
 b_t = D \frac{\alpha + \| \mbx_t - \mbx_o\|}{\| \mbx_t - \mbx_o\|}. \nn
\end{equation}
Thus (\ref{E:Integral_approx3}) simplifies to become
\begin{eqnarray} \label{E:Integral_approx4}
 Q &=& Q_t + Q_r, \quad \mbox{where}
\\
 Q_t &=&  \int_{\alpha=0}^{\infty} \eta D \frac{\alpha + \| \mbx_t - \mbx_o\|}{\| \mbx_t - \mbx_o\|} f\left(\mbx_o + \alpha \frac{\mbx_o - \mbx_t}{\| \mbx_o - \mbx_t \|} \right) d\alpha
\nnn
 Q_r &=&  \int_{\alpha=0}^{\infty} \eta D \frac{\alpha + \| \mbx_r - \mbx_o\|}{\| \mbx_r - \mbx_o\|} f\left(\mbx_o + \alpha \frac{\mbx_o - \mbx_r}{\| \mbx_o - \mbx_r \|} \right) d\alpha
\nn
\end{eqnarray}

\subsection{ETAP: Scattering} \label{S:totalScatterPower}

First, we consider the expected total affected power (ETAP) assuming that all multipath are due to scattering. We use the scattering power formula of (\ref{E:Norklit}) as the function $f(\mbx)$ in (\ref{E:Integral_approx4}). 

We note that there is a singularity in (\ref{E:Norklit}) whenever the scatter position $\mbx$ is equal to the TX or RX position.  Infinite received power is not physically possible -- it is an artifact of a far-field approximation formula evaluated at near-field -- thus it is appropriate to ensure that $P_s(\mbx)$ is finite for all positions $\mbx$.  The singularity can be remedied by adding a constant into the denominator, as studied in \cite{inaltekin09}.  Instead, we have specified that the scatterer plane does not contain the TX or RX locations, thus avoiding the singularity.

We apply $f(\mbx) = P_s(\mbx)$ to find $Q = Q_t + Q_r$.  For $Q_t$, the integral in (\ref{E:Integral_approx4}) simplifies to,
\begin{equation} \label{E:Qt}
Q_t =  \int_{\alpha=0}^{\infty}  \frac{D \eta c_s / \|\mbx_t-\mbx_o\|}{\left( \| \mbx_t - \mbx_o\| + \alpha \right) 
       \left\| (\mbx_r - \mbx_o) - \alpha \frac{\mbx_o-\mbx_t}{\|\mbx_o-\mbx_t\|} \right\|^2} d\alpha
\end{equation}
The expression for $Q_r$ is obtained by switching $\mbx_t$ and $\mbx_r$ in (\ref{E:Qt}). 
We show in the Appendix that $Q$ simplifies to
\begin{equation} \label{E:FinalEPaff}
%  Q = \frac{Dc\eta}{\| \mbx_r - \mbx_t\|^2} 
%                   \left\{  
%                       \frac{(\pi - \theta)(1+\cos \theta)}{d^{+}_{\mbx_o} \sin \theta} 
%                    + \right. \nnn
%                 \left. \frac{1}{d^{-}_{\mbx_o}}\log \frac{\| \mbx_t - \mbx_o\|}{\| \mbx_r - \mbx_o\|} 
%                   \right\}
 Q = \frac{Dc_s \eta}{d_{rt}^2} 
                  \left\{
                      \frac{(\pi - \theta)(1+\cos \theta)}{d^{+}_{\mbx_o} \sin \theta} 
                   + \frac{1}{d^{-}_{\mbx_o}} \log \frac{\| \mbx_t - \mbx_o\|}{\| \mbx_r - \mbx_o\|}
                  \right\}
\end{equation}
where $d_{rt} = \| \mbx_r - \mbx_t\|$, and
\begin{eqnarray} \label{E:theta}
   d^{+}_{\mbx_o} &=& (\| \mbx_r - \mbx_o\|^{-1} +  \| \mbx_t - \mbx_o\|^{-1} )^{-1}, \nnn 
   d^{-}_{\mbx_o} &=& (\| \mbx_r - \mbx_o\|^{-1} -  \| \mbx_t - \mbx_o\|^{-1} )^{-1}, \nnn
 \theta &=& \cos^{-1}\left( \frac{(\mbx_r - \mbx_o)^T(\mbx_o - \mbx_t)} {\| \mbx_r - \mbx_o\| \| \mbx_o - \mbx_t\|} \right).
\end{eqnarray}
Note that $d^{+}_{\mbx_o}$ is the combination in parallel of the two distances from the TX and RX to the person, and $d^{-}_{\mbx_o}$ is the parallel difference of the two distances. The angle $\theta$ is the angle between the line from $\mbx_o$ to $\mbx_r$ and the line from $\mbx_t$ to $\mbx_o$.  Note that because of the $\cos^{-1}(\cdot)$, $0 \le \theta \le \pi$.  

\begin{figure}[tb]
  \centerline{(a) \psfig{figure=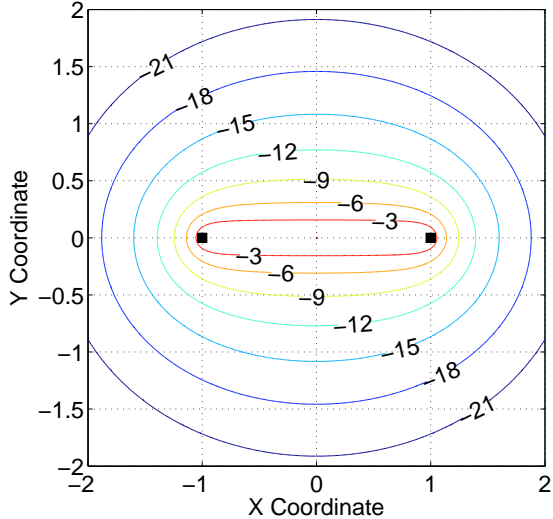,width=2.9in}}
  \centerline{(b) \psfig{figure=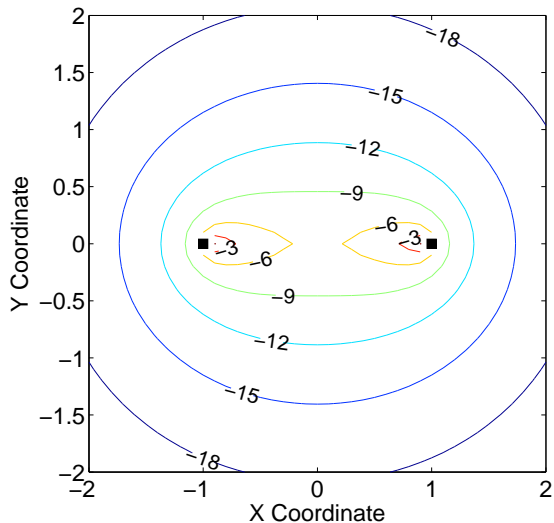,width=2.9in}}
  \caption{ETAP in dB relative to the maximum, with $\mbx_t$ and $\mbx_r$ as shown ($\blacksquare$) and $\Delta z = 0.1$m, for (a) scattering, and (b) reflection (with $n_p=3$).}
  \label{F:plot_log_exp_aff_power_contour}
\end{figure}

For the case of $\mbx_t=[-1, 0, 0.1]^T$ and $\mbx_r=[1, 0, 0.1]^T$, the ETAP as a function of $\mbx_o$ is shown in Figure \ref{F:plot_log_exp_aff_power_contour}(a).  Note that since the plot is shown in dB relative to the maximum value, the constants $D$, $c_s$, and $\eta$ do not change the plot.

\subsection{ETAP: Reflection} \label{S:totalScatterPower}

In this section we calculate the ETAP assuming that reflection is the only propagation mechanism.  In this case, the function $f(\mbx)$ is set equal to $P_r(\mbx)$ as given in (\ref{E:Liberti2}).  We still have that $Q=Q_t + Q_r$, and the integral expression for $Q_t$ is
\begin{eqnarray} \label{E:Reflection_ETAP}
  Q_t =  \int_{\alpha=0}^{\infty} \frac{\eta D c_r \frac{\alpha + \| \mbx_t - \mbx_o\|}{\| \mbx_t - \mbx_o\|}  d\alpha }
{ \left( \| \mbx_t - \mbx_o \| + \alpha + \left\| \mbx_r - \mbx_o - \alpha \frac{\mbx_o - \mbx_t}{\| \mbx_o - \mbx_t \|} \right\| \right)^{n_p} } 
\end{eqnarray}
Unfortunately, this $Q_t$ is not generally tractable for arbitrary real-valued path loss exponents $n_p$.   Instead, we perform numerical integration to find $Q_t$ and $Q_r$.  Compared to the exact equation for reflection ETAP, the expression in (\ref{E:Reflection_ETAP}) does simplify calculation by requiring only one integral.  For the case of $n_p = 3$, $\mbx_t=[-1, 0, 0.1]^T$ and $\mbx_r=[1, 0, 0.1]^T$, the ETAP as a function of $\mbx_o$ is shown in Figure \ref{F:plot_log_exp_aff_power_contour}(b). Again, since the plot is in dB relative to the maximum, the constants $D$, $c_r$, and $\eta$ do not change the plot.

\subsection{Parameter Sensitivity}

In this section, we study the effects on ETAP of changes in path loss exponent, $n_p$, and the relative height of the plane containing the TX and RX, $\Delta z$.
%How does the power along the line between the RX and TX vary with parameter changes?  Particularly, $n_p$ and distance between the TX/RX plane and the scattering plane.  This section will present results for these two parameters.

\begin{figure}[htb]
  \centerline{ \psfig{figure=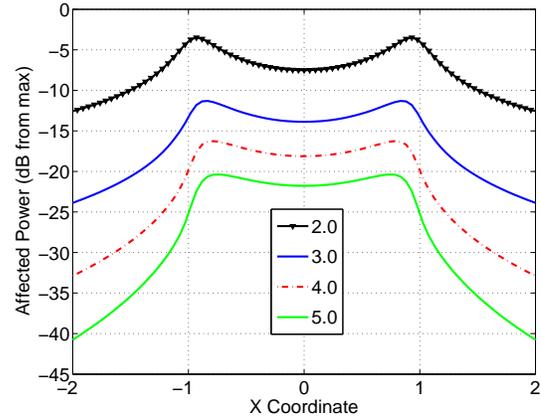,width=2.8in}}
  \caption{Reflection ETAP (dB max) for $\mbx_o$ with y-coordinate equal to 0.1, for $n_p \in \{2.0, 3.0, 4.0, 5.0\}$, and $\mbx_t=[-1, 0, 0.1]^T$ and $\mbx_r=[1, 0, 0.1]^T$.}
  \label{F:plot_ETAP_reflection_many_np}
\end{figure}

For the reflection ETAP, we vary the path loss exponent $n_p$.  We show the results by plotting in Figure \ref{F:plot_ETAP_reflection_many_np} a horizontal cut of the value of ETAP (along the line with y-coordinate equal to 0.1).  Note that all plots are in dB relative to the maximum of ETAP across all space and for all tested values of $n_p$.  We observe that increasing path loss exponent generally decreases the ETAP.  Further, away from the TX and the RX (below -1 and above +1), an increasing $n_p$ also increases the rate of decrease in ETAP.  In effect, the area with high ETAP is more isolated from the area with low ETAP when $n_p$ is higher.  So although higher $n_p$ reduces the magnitude of the ETAP, it may benefit radio tomography by providing greater location precision when ETAP (or variance of RSS) is measured to be high.

\begin{figure}[htb]
  \centerline{ (a) \psfig{figure=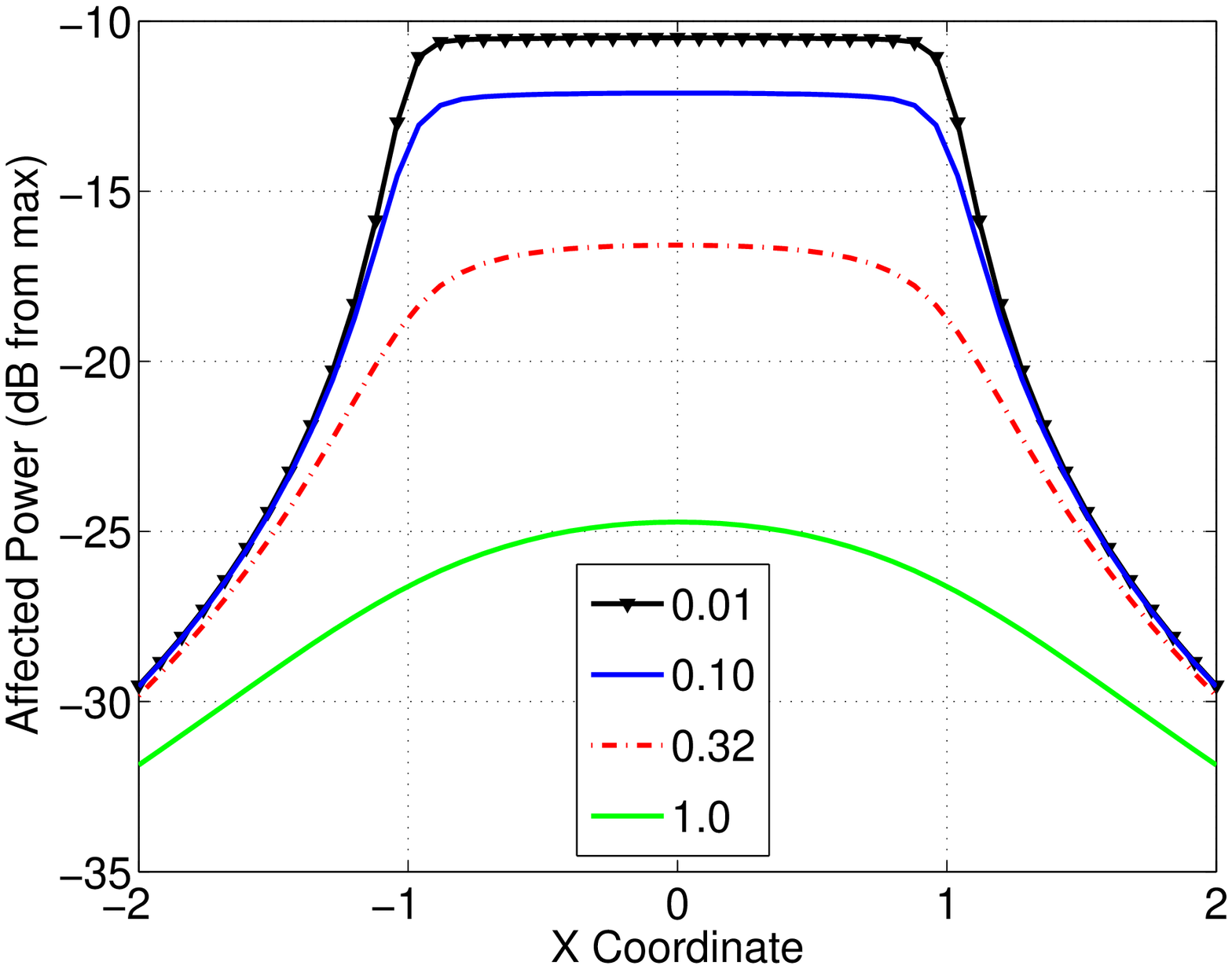,width=2.8in}} 
  \centerline{ (b) \psfig{figure=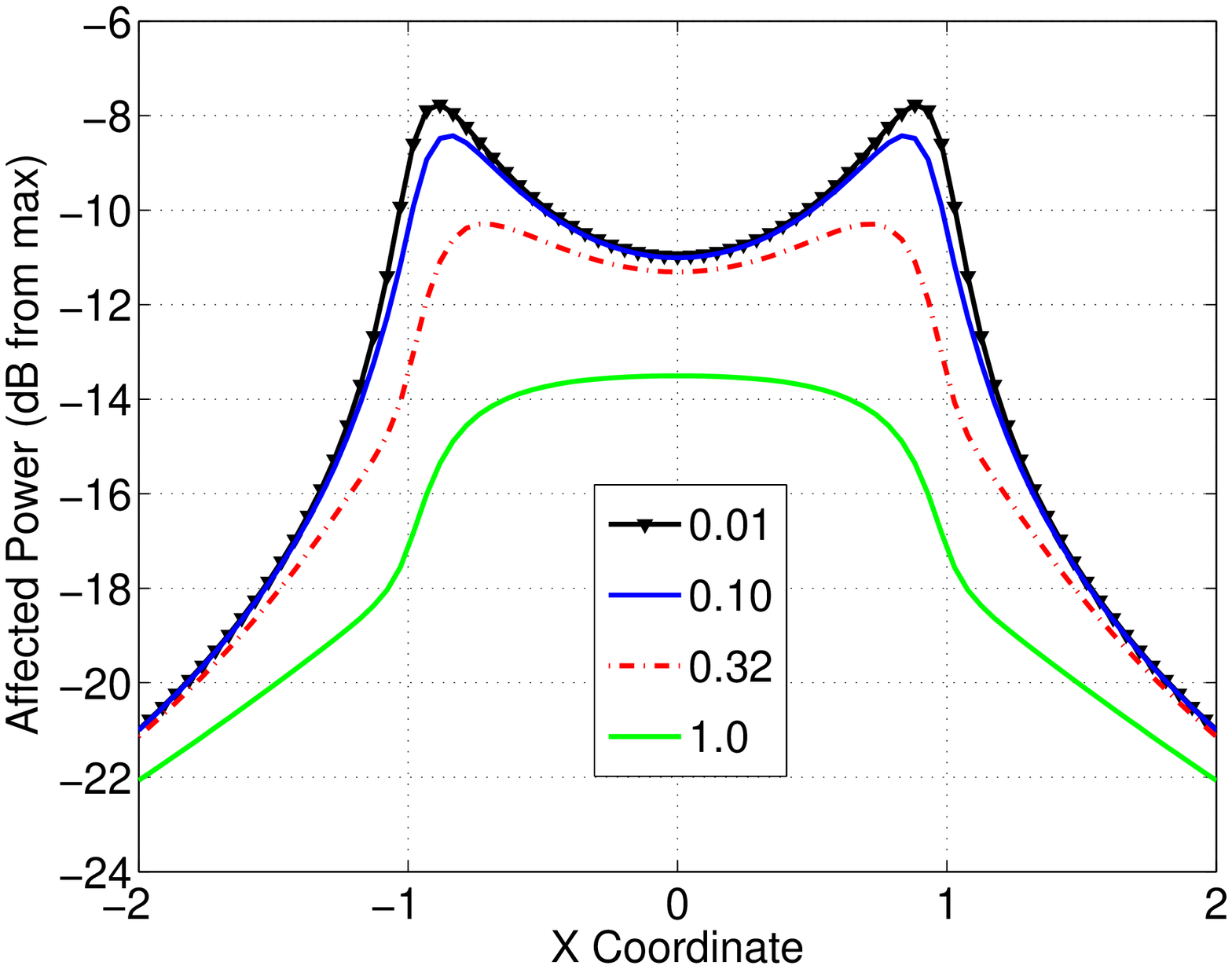,width=2.8in}} 
  \caption{ETAP (dB max) vs.~$\mbx_o$ with y-coordinate equal to 0.1, for several $\Delta z$, when $\mbx_t=[-1, 0, \Delta z]^T$ and $\mbx_r=[1, 0, \Delta z]^T$, for (a) reflection and (b) scattering.}
  \label{F:plot_ETAP_reflection_many_planeHeight}
\end{figure}

Next, we consider the effect of the separation between the TX/RX plane and scatterer plane, $\Delta z$.  For the case of reflection, the effects are shown in Figure \ref{F:plot_ETAP_reflection_many_planeHeight}(b).  For low values of $\Delta z$, when the scattering plane contains or nearly contains the TX and RX, there are noticeable peaks in ETAP near the TX and RX.  In contrast, when the TX and RX are in a plane high above (or low below) the scatterer plane, the shape of the ETAP surface actually turns around such that its maximum is halfway between the TX and RX.

For the case of scattering, the effects are shown in Figure \ref{F:plot_ETAP_reflection_many_planeHeight}(a).  In the scattering case, the ETAP is always highest halfway between the TX and RX.  But, with increasing $\Delta z$ the shape of the ETAP becomes rounder, and the slope outside of the area between the TX and RX becomes shallower.  

\section{Discussion} \label{S:AnalysisDiscussion}

A few basic assumptions and approximations have allowed analysis of affected multipath, as a function of $\mbx_o$ and the TX and RX positions.  These quantities include the total number of affected multipath and the ETAP when propagation is due to scattering or reflection.  Previous research has surmised that the new multipath caused by scattering from the person was the sole change in the channel; and that this change introduced a new path with power given by the bistatic scattering equation in (\ref{E:Norklit}).  The contours of this power surface $P_s(\mbx_o)$ have the Cassini oval shape.  

The ETAP expressions in this section, in combination with the ETAP/RSS variance relationship presented in Section \ref{S:Variance_ETAP}, show the ``shape'' of the typical RSS variance as a function of person position $\mbx_o$.  By considering ETAP, in contrast to past research, we have shown that reflection, not scattering, leads to RSS variance surfaces with contours similar in shape to Cassini ovals.  The reflection ETAP has two peaks around the TX and RX positions, except when $\Delta z$ is large.  In contrast, Figure \ref{F:plot_log_exp_aff_power_contour}(a), which shows the scattering ETAP, does not show peaks around the TX and RX positions for any $\Delta z$.  Instead, Figures \ref{F:plot_log_exp_aff_power_contour}(a) and \ref{F:plot_ETAP_reflection_many_planeHeight}(a) show a single rounded rectangular peak when $\mbx_o$ is between the TX and RX locations.

In summary, when reflection is the dominant mechanism, and the scattering plane is close to the plane containing the transceivers, the maximum effect is seen when the person is very close to either the RX or TX.  When scattering is the dominant mechanism, or when the scattering plane is far from the plane containing the transceivers, the maximum effect is seen when the person is halfway between the RX and TX.  

\section{Experimental Results} \label{S:ExperimentalResults}

We refer in the Introduction to two studies which reported experiments which quantify the RSS variation as a function of the TX, RX, and person's location.  The results of \cite{yao2008model}, shown in Figure \ref{F:yao-gao-figure4}, show highest variation when the person is close to either node and high variation when the person is close to the line connecting the nodes.  In contrast, experimental results in \cite[Figs.~4, 5]{zhang2007rf} show a different result, that the ``dynamic'' (average absolute value of the difference from the static mean) is highest in an oval centered at the midpoint of the line segment between the two nodes.  

The differences in the experimental setups may explain the different results.  In \cite{yao2008model}, RF sensors are located on tripods, at an apparent height of about 1.5 meters, thus the plane containing the TX and RX is likely to also include many scatterers.  Since $\Delta z$ is close to zero the analysis would tell us to expect the highest lines from Figures \ref{F:plot_ETAP_reflection_many_planeHeight}(a) and (b). Because the variance is highest closest to the TX and RX, the analysis then suggests that the main propagation mechanism is reflection.

In contrast, in \cite{zhang2007rf}, all RF sensors are located on the ceiling (at 2.4 meters) in an empty room. Because the room is empty, the scattering plane (likely the floor) is separated by $\Delta z = 2.4$ meters from the plane containing the TX and RX.  Neighboring sensors are separated by 2 meters. Since $\Delta z$ is greater than the distance between sensors, the analysis indicates to expect the lowest lines in Figures \ref{F:plot_ETAP_reflection_many_planeHeight}(a) and (b).  Both reflection and variance ETAP at high $\Delta z$ have a shape which is highest in the midpoint between the TX and RX, similar to the experimentally reported results. 

\begin{figure}[tb]
  \centerline{ \psfig{figure=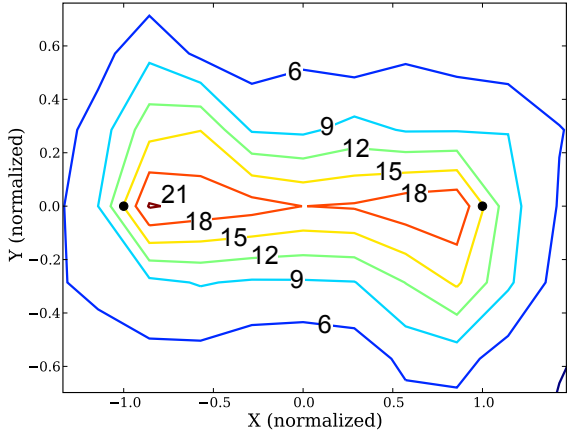,width=3.4in}}
  \caption{Experimental RSS variance vs.~$\mbx_o$, normalized such that the RX is located at (-1, 0) and the TX is located at (1,0).}
  \label{F:Joeys_Var}
\end{figure}

We perform an extensive experiment using the following procedure to experimentally determine the RSS variance surface in another environment.  Thirty-four nodes (Crossbow TelosB, operating the IEEE 802.15.4 protocol at 2.4 GHz) are placed throughout a 16 by 36 feet area within the University of Utah Bookstore, all at approximately 1.2 meter height.  It is a typical store environment, with bookshelves, books, aisles, tables, chairs, product displays, and other obstructing objects at height similar to the node height.  The location of each node is surveyed and recorded.  The signal strength of each link is recorded over time as a human walks in a known track within the area.  For each measurement of RSS, $R_{dB}$, on each link, the coordinates of the TX, RX, and human are all known.  We rotate, translate, and scale the coordinates such that $\mbx_t = [1, 0]^T$ and $\mbx_r=[-1, 0]^T$, that is, we normalize the coordinates so that the TX and RX coordinates are the same as used in the examples in Figure  \ref{F:plot_log_exp_aff_power_contour}.  The person's position is transformed in the same manner to obtain $\mbx_o$ so that its relative position w.r.t. $\mbx_t$ and $\mbx_r$ is preserved.   In post-processing, we create a two dimensional grid of bins across the range of $\mbx_o$, and each RSSI measurement $R_{dB}$ for each time sample is stored in a list for the bin that corresponds to the human position $\mbx_o$.  This process is repeated for every link measurement made during the experiment, over 110,000 measurements.  The variance of RSS measurements for each bin is calculated and plotted as a surface.  A contour plot of the variance surface is given in Figure \ref{F:Joeys_Var}.

Because in our experiments the scattering plane is similar in height to the plane containing the RF sensors, we have $\Delta z$ close to zero.  Similar to \cite{yao2008model}, we by comparing the results in Figure \ref{F:Joeys_Var} to the ETAP analysis that the dominant propagation mechanism was reflection.

% \section{Accuracy of Approximations}
% 
% This section will remove Approximations 1, 2, and 3 and calculate the double integral for $Q$ and compare the results.

\section{Conclusion}

This paper presents models for the variance of RSS caused by motion of a person (or object), as a function of the person's position, $\mbx_o$, and the TX and RX positions, $\mbx_t$ and $\mbx_r$.  The analysis is statistical, rather than based on any specific knowledge of the environment, and considers single-bounce multipath caused by scattering and reflection.  We expect to see surfaces with contours of Cassini ovals when considering scattering from the person.  However, this new analysis shows two interesting things:
\begin{itemize}
 \item Even when multipath are all caused by reflection, variance surfaces vs.~$\mbx_o$ can have contours similar to Cassini ovals.
 \item When multipath are caused by scattering, or in high $\Delta z$ cases of reflection, variance surfaces are highest when the person is in the middle of the line between the TX and RX.
\end{itemize}
We show using reported measurements, and a new extensive measurement set, that the analysis shows results that are similar to those observed experimentally.

Future analytical and experimental work is required to verify and extend the results presented in this paper.  Analytical results should be extended to the case of multiple `bounces', \ie, when multipath may experience more than one reflection or scattering.  Experimental work should measure the total affected multipath power in a variety of controlled and uncontrolled environments to test the relationships derived in this paper.  Finally, we will apply the developed statistical model to improve estimation bounds and imaging and tracking algorithms for use in DFL systems.

%By populating a controlled environment with ``rough'' surfaces (which are scatterers at the center frequency) we would expect to see more scattering within the channel, and thus should experience variance with a shape more similar to Figure \ref{F:plot_log_exp_aff_power_contour} than to Figure \ref{F:plot_log_exp_TAP_ReflModel_contour}.

\appendix

\section{Proof of Scattering ETAP} \label{S:Proof}

In this section, we prove (\ref{E:FinalEPaff}) from (\ref{E:Qt}).  Expanding from (\ref{E:Qt}),
\begin{eqnarray} \label{E:Vt2}
Q_t &=&  
\frac{Dc_s \eta}{ \|\mbx_t - \mbx_o \|} \int_{\alpha=0}^{\infty}  \frac{d \alpha}{ ( a + \alpha ) ( \alpha^2 - f \alpha  +  b^2)} \nn
\end{eqnarray}
where we define $a = \| \mbx_t - \mbx_o\|$, $b=\|\mbx_r-\mbx_o\|$, and $f = 2 \|\mbx_r-\mbx_o\| \cos \theta$.  By partial fraction expansion, the fraction in the integral can be written as,
\[
 \frac{1}{a^2 + af + b^2} \left[
          \frac{1}{a + \alpha} + \frac{-\frac{1}{2}(2\alpha - f)  + \frac{1}{2}(2a + f)}{\alpha^2 - f \alpha  +  b^2} 
   \right]
\]
Note that
\begin{eqnarray}
 a^2 + fa + b^2 &=& 2 \|\mbx_r-\mbx_o\| \| \mbx_t - \mbx_o\| \cos \theta  \nnn
               & & + \| \mbx_t - \mbx_o\|^2 + \|\mbx_r-\mbx_o\|^2 
 \nnn
                &=& \| (\mbx_o - \mbx_t) + (\mbx_r-\mbx_o)\|^2 = \| \mbx_r - \mbx_t\|^2 \nn
\end{eqnarray}
Using an integral table \cite[2.103(4)]{gradshteyn} we can obtain,
\begin{eqnarray}
 Q_t &=&  \frac{Dc_s \eta}{\|\mbx_t-\mbx_o\| \| \mbx_r - \mbx_t\|^2}  \left[ A +  B\right|_{\alpha=0}^\infty \nn
\end{eqnarray}
where 
\begin{eqnarray}
A &=&    \frac{2a+f}{\sqrt{4b^2-f^2}} \arctan \frac{2\alpha-f}{\sqrt{4b^2-f^2}}, \nnn
B &=& \log \frac{a+\alpha}{\sqrt{\alpha^2 - f\alpha + b^2}}, \nn
\end{eqnarray}
where $\arctan(x)$ is the arctangent of $x$.   Note that $\left[ B \right|_{\alpha=0}^\infty = \log 1 - \log\frac{a}{b}$.
For the $A$ term, we have that $\sqrt{4b^2 -f^2} = 2 \|\mbx_r-\mbx_o\|\sqrt{1-\cos^2 \theta} = 2 \|\mbx_r-\mbx_o\|\sin \theta$, where the absolute value around the sine term is not necessary since $0 \le \theta \le \pi$.   Finally, $2a+f = 2(\|\mbx_t-\mbx_o\| + \|\mbx_r-\mbx_o\|\cos \theta)$.   
Evaluating the limits on $A$, 
\begin{eqnarray}
 \left[ A \right|_{\alpha=0}^\infty &=& \frac{\pi  - \arctan(\cot \theta)}{\sin \theta}  \left[
                              \frac{\|\mbx_t-\mbx_o\|}{\|\mbx_r-\mbx_o\|} + \cos\theta
                          \right]  \nn
\end{eqnarray}
And  thus expressions for $Q_t$, and similarly, for $Q_r$, are:
\begin{eqnarray}
 Q_t &=& \frac{Dc_s \eta}{\| \mbx_r - \mbx_t\|^2} \left\{  
                          \frac{(\pi - \theta)}{\sin \theta} \left[
                              \frac{1}{\|\mbx_r-\mbx_o\|} + \frac{\cos\theta}{\|\mbx_t-\mbx_o\|}
                          \right]  \right. \nnn
 &&
    \left.              - \frac{1}{\|\mbx_t-\mbx_o\|}\log\frac{\| \mbx_t - \mbx_o\|}{\| \mbx_r - \mbx_o\|} 
                 \right\}    \nnn
 Q_r &=& \frac{Dc_s \eta}{\| \mbx_r - \mbx_t\|^2} \left\{  
                          \frac{(\pi - \theta)}{\sin \theta} \left[
                              \frac{1}{\|\mbx_t-\mbx_o\|} + \frac{\cos\theta}{\|\mbx_r-\mbx_o\|}
                          \right]  \right. \nnn
 &&
    \left.              - \frac{1}{\|\mbx_r-\mbx_o\|}\log\frac{\| \mbx_r - \mbx_o\|}{\| \mbx_t - \mbx_o\|} 
                 \right\}    
\end{eqnarray}
Plugging the final expressions for $Q_t$ and $Q_r$ into (\ref{E:Integral_approx4}), we have the expression in (\ref{E:FinalEPaff}).

\section{Approx.~Expected Logarithm} \label{S:ApproxExpectedLog}

In this section, we justify that $\E{}{10 \log_{10} Y} \approx 10\log_{10} \E{}{ Y}$, for $Y = \sum_{i\in \mathcal{T}'} |V_i |^2$.  Converting to natural logarithms and representing $|V_i |^2 = V_{i,I}^2 + V_{i,Q}^2$, where $V_{i,I}$ and $V_{i,Q}$ are the real and imaginary components of $V_i$,
\begin{equation}\label{E:appendix_start}
 \E{}{10 \log_{10} Y} = \frac{10}{\log(10)} \E{}{\log \sum_{i\in \mathcal{T}'} V_{i,I}^2 + V_{i,Q}^2}
\end{equation}

We can't evaluate (\ref{E:appendix_start}) without making a distributional assumption about $V_{i,I}$ and $V_{i,Q}$.  It is typically assumed that $\{V_{i}\}_i$ are zero-mean complex Gaussian random variables \cite{durgin02}.  In our case, we assume that $V_{i,I}$ and $V_{i,Q}$ are i.i.d.~with variance $\sigma^2$.  Clearly, not all multipath components have the same variance, since multipath caused by more distant scatterers tend to have lower power.  However, the equal variance case is one that is analytically tractable and can be used to show the accuracy of the approximation.   

From (\ref{E:appendix_start}), we renumber the multipath as 1 through 2m, w.l.o.g,  where $m=|\mathcal{T}'|$,
\begin{equation}\label{E:E_log_sum_2m}
 \E{}{10 \log_{10} Y} = \frac{10}{\log(10)} \E{}{\log \sum_{j=1}^{2m} |U_j|^2}
\end{equation}
where $U_j$ are i.i.d. Gaussian with zero mean and variance $\sigma^2$.  Moser \cite{moser08} has shown that the expected value of the logarithm of this sum evaluates to
\begin{equation}\label{E:E_log_sum_Moser}
 \E{}{10 \log_{10} Y} = \frac{10 }{\log(10)} \left[\left( -\gamma_E + \sum_{j=1}^{2m-1} \frac{1}{j} \right)\sigma^2 \right]
\end{equation}
where $\gamma_E$ is Euler's constant, $\gamma_E \approx 0.577$.  In this case, our approximation would be
\begin{equation}\label{E:E_log_sum_2m_approx}
 10 \log_{10} \E{}{Y} = 10 \log_{10} \sum_{j=1}^{2m} \E{}{|U_j|^2} = 10 \log_{10} 2m \sigma^2
\end{equation}
We compare (\ref{E:E_log_sum_2m_approx}) and (\ref{E:E_log_sum_Moser}) numerically.  For one multipath, the worst case, the approximation is in error by 1.2 dB.  With two multipath, the error is 0.6 dB, and with five or more multipath, the error is 0.2 dB or less.  

% \begin{figure}[htbp]
%   \centerline{ \psfig{figure=plotExpectedLog.eps,width=2.7in}}
%   \caption{A comparison of the exact expression for $\E{}{10 \log_{10} Y}$ vs. the approximation, $ 10 \log_{10} \E{}{Y}$, for $m$ i.i.d.~multipath components.}
%   \label{F:plotExpectedLog}
% \end{figure}

In summary, for the case of i.i.d.~zero-mean complex Gaussian multipath, the error in approximating $\E{}{10 \log_{10} Y}$ as $10\log_{10} \E{}{ Y}$ is small even for small numbers of multipath.  As more multipath are present, the approximation becomes more exact.

%\bibliographystyle{abbrv}
%\bibliography{overall}

\end{document}

%% file: header.tex
\newcommand{\Var}[2]{\mbox{Var}_{#1}\left[ {#2} \right]}

\newcommand{\E}[2]{E_{#1}\left[ {#2} \right]}

\newcommand{\ie}{{\it i.e.}}

\newcommand{\mbx}[0]{{\mathbf{x}}}

\newcommand{\nnn}[0]{\nonumber \\ }
\newcommand{\nn}[0]{\nonumber }